\documentclass[prd,twocolumn,showpacs,floatfix,amsmath,nofootinbib,amssymb,floatfix]{revtex4}
\usepackage{graphicx,color,dcolumn,booktabs,bm,multirow}
\usepackage{longtable,lscape}
\usepackage{txfonts}
 \usepackage{enumitem}
\usepackage{overpic}
\usepackage{amssymb}
\usepackage{indentfirst}
\usepackage{feynmf}   %{feynmp}
\usepackage{slashed}  %for Feynman symbols
\usepackage{cases}
\usepackage{color}
\usepackage{multirow}
\usepackage{epstopdf}
\usepackage{float}
\usepackage{graphicx,color,dcolumn,booktabs,bm}

\usepackage[colorlinks,
            citecolor=blue,
            anchorcolor=red,
            menucolor=red,
            linkcolor=red,
            filecolor=red,
            runcolor=red,
            urlcolor=blue,
            frenchlinks=red]{hyperref}
\begin{document}

\title{The $X(3872)\rightarrow J/\psi \pi \gamma$ and $X(3872)\rightarrow J/\psi \pi\pi \gamma$ decays}
\author{Qi Wu}\email{wu\_qi@pku.edu.cn}
\author{Jun-Zhang Wang}\email{wangjzh2022@pku.edu.cn}
\author{Shi-Lin Zhu}\email{zhusl@pku.edu.cn}
\affiliation{School of Physics and Center of High Energy Physics,
Peking University, Beijing 100871, China}

\begin{abstract}
We study the $\rho$ and $\omega$ meson contribution to the radiative
decays $X(3872)\rightarrow J/\psi \pi \gamma$ and
$X(3872)\rightarrow J/\psi \pi\pi \gamma$. The $X(3872)\rightarrow
J/\psi \pi \gamma$ is dominated by the $\omega$ meson. As for the
$X(3872)\rightarrow J/\psi \pi\pi \gamma$, the contributions of the
cascade decays through the $\rho$ and $\omega$ mesons are strongly
suppressed with respect to the diagrams which proceed either through
the $\psi(2S)$ or the three body decay of $\rho$. The branching
ratios of $X(3872)\rightarrow J/\psi \pi \gamma$ and
$X(3872)\rightarrow J/\psi \pi\pi \gamma$ are
$(8.10^{+3.50}_{-2.88})\times10^{-3}$ and $(2.38\pm1.06)\%$, which
may be accessible by the BESIII and LHCb Collaborations. Especailly,
the $X(3872)\rightarrow J/\psi \pi \gamma$ and $X(3872)\rightarrow
J/\psi \pi^+\pi^- \gamma$ decays can be employed to extract the
couplings $g_{X\psi\omega}$ and $g_{X\psi\rho}$, which probe the
isoscalar and isovector components of the X(3872) wave function
respectively.

\end{abstract}

\date{\today}
 \maketitle

%%%%%%%%%%%%%%%%%%%%%%%%%%%%%%%%%%
\section{Introduction}
\label{sec:introduction}

Twenty years ago, a new narrow charmonium-like state $X(3872)$ was
observed in the exclusive decay process $B^+ \to K^+ \pi^+ \pi^-
J/\psi$ by the Belle Collaboration~\cite{Belle:2003nnu}, which
opened a door to the exotic hadron spectroscopy (see
Refs.~\cite{Chen:2016qju,Hosaka:2016pey,Lebed:2016hpi,Esposito:2016noz,Guo:2017jvc,Ali:2017jda,Olsen:2017bmm,Karliner:2017qhf,Yuan:2018inv,Dong:2017gaw,Liu:2019zoy,
Chen:2022asf,Meng:2022ozq} for recent reviews). After its discovery,
the $X(3872)$ was subsequently confirmed by several other
experiments~\cite{CDF:2003cab,D0:2004zmu,BaBar:2004oro}. Ten years
after its discovery, its spin-parity quantum numbers were finally
determined to be $J^{PC}=1^{++}$ by the LHCb
Collaboration~\cite{LHCb:2015jfc}. None of the charged partner of
the $X(3872)$ was found ~\cite{BaBar:2004cah}.

Since the discovery of the $X(3872)$, there have been tremendous
efforts to investigate its inner structure experimentally and
theoretically. The exotic nature of the $X(3872)$ was embodied in
its mass and width, which are listed in Table
~\ref{Tab:X_parameters}.
\begin{table}[htp]
 \centering
 \caption{The resonance parameters of the $X(3872)$ from Particle Data Group (PDG)~\cite{Workman:2022ynf} (in units of MeV), where $\Delta E=M_X-M_{Threshold}$ . \label{Tab:X_parameters}}
 \begin{tabular}{p{2cm}<\centering p{2cm}<\centering p{2cm} <\centering p{2cm} <\centering p{2cm}}
 \toprule[1pt]
  Mass  & Width & Threshold & $\Delta E$ \\
 %\midrule[1pt]
  $3871.65 \pm 0.06$  & $1.19 \pm 0.21$ & $D^+ D^{*-}/D^0 \bar{D}^{*0}$ & -8/-0.04\\
 \bottomrule[1pt]
 \end{tabular}
\end{table}
One of the most intriguing feature of the $X(3872)$ is that its mass
almost coincides with the $D^0 \bar{D}^{*0}$ threshold. Considering
the narrow width of $X(3872)$, it is natural to regard the $X(3872)$
as a $D \bar{D}^{*}$ hadronic
molecule~\cite{Voloshin:2003nt,Swanson:2003tb,Tornqvist:2004qy,Fleming:2007rp,Liu:2008fh,Tornqvist:1993ng}.
The molecule picture not only explains the coincidence of the mass
of the $X(3872)$ with the $D^0 \bar{D}^{*0}$ threshold naturally,
but also explains its isospin violation in the $J/\psi\rho$ decay
mode~\cite{Swanson:2003tb,Swanson:2004pp,Li:2012cs}. However, some
molecule models meet with difficulties when explaining the following
phenomena:
\begin{itemize}

\item The ratio $\Gamma(B^0\to K^0 X)/\Gamma(B^+\to K^+ X)$ is about unity according to an estimation based on the molecule picture~\cite{Braaten:2007dw,Braaten:2007ft,Braaten:2003he}, which is about two times larger than measurements by the BaBar~\cite{BaBar:2004oro} and Belle~\cite{Belle:2011vlx} Collaborations.
\item The predicted branching ratios of $X(3872)\to D^0 \bar{D}^0 \gamma$ and $X(3872)\to J/\psi \gamma$~\cite{Swanson:2003tb,Swanson:2004pp} largely deviated from the experimental data.
\item As a loosely bound hadronic molecule with a small binding energy, $X(3872)$ was expected to be so fragile that it would be hard to explain the observed production rate in the high energy $p\bar{p}$ collisions at the Tevatron~\cite{Bignamini:2009sk}.
\end{itemize}
Actually, the above difficulties indicate that there should exist a
significant $c\bar{c}$ component in the wave function of the
$X(3872)$~\cite{Suzuki:2005ha,Li:2009zu}. In other words, the
coupled channel effect may play an important role in the formation
of the $X(3872)$.

To date, the inner structure of the $X(3872)$ is still an open
question and remains challenging. In addition to the mass spectrum,
the decay patterns also encode important dynamical information and
hence provide another perspective about its underlying structure.
The ratio $\mathcal{B}[X\to J/\psi \pi^+\pi^-\pi^0]/\mathcal{B}[X\to
J/\psi \pi^+\pi^-]$ has been measured by several
experiments~\cite{Abe:2005ix,delAmoSanchez:2010jr,Ablikim:2019zio},
which indicates a large isospin violation. This ratio is of great
interest and has been investigated in different
scenarios~\cite{Tornqvist:2004qy,Suzuki:2005ha,Ortega:2009hj,Gamermann:2009fv,Hanhart:2011tn,Li:2012cs,Zhou:2017txt,Wu:2021udi,Meng:2021kmi}.
Different components in the wave function of the $X(3872)$ will
affect the decays either in the long distance or the short distance.
In other words, the decay patterns encode very important information
on the underlying structure and can be used to test different
theoretical explanations. For example, the $X\to D^0 \bar{D}^0
\pi/\gamma$, which proceeds through the decays of either $D^{*0}$ or
$\bar{D}^{*0}$ and thus belongs to the long-distance decays, can be
used to study the long-distance structure of the
$X(3872)$~\cite{Voloshin:2003nt}. Pionic transitions from the
$X(3872)$ to $\chi_{cJ}$ were investigated in
Refs.~\cite{Dubynskiy:2007tj,Fleming:2008yn,Mehen:2015efa}. The
relative rates for these transitions to the final states with
different $J$ is very sensitive to the inner structure of the
$X(3872)$ as a pure charmonium state or a four-quark/molecular state
~\cite{Dubynskiy:2007tj}. The predictions of the ratio
$\mathcal{B}[X\to \psi^\prime \gamma]/\mathcal{B}[X\to J/\psi
\gamma]$ from the $D\bar{D}^*$
molecule~\cite{Swanson:2003tb,Ferretti:2014xqa}, pure charmonium
state~\cite{Barnes:2005pb} and molecule$-$charmonium
mixture~\cite{Badalian:2012jz,Dong:2009uf} turned out to be
dramatically different from each other, which reflects the
importance of the $c\bar{c}$ component in the $X(3872)$.

\begin{table}[htb]
\centering
\renewcommand{\arraystretch}{1.2}
    \caption{The branching ratios ($\%$) of $X(3872)$ from PDG~\cite{Workman:2022ynf}.}\label{tab:br}
    \setlength{\tabcolsep}{6.mm}
\begin{tabular}{ccc}
  \toprule[1pt]\toprule[1pt]
  Decay channels & Branching ratios \\
  \midrule[1pt]
  $\pi^+ \pi^- J/\psi$ & $3.8\pm1.2$  \\
  $\omega J/\psi$ & $4.3\pm2.1$ \\
  $D^0 \bar{D}^0 \pi^0$ & $49^{+18}_{-20}$ \\
 $D^0 \bar{D}^{*0}$ & $37\pm9$ \\
 $\pi^0 \chi_{c1}$ & $3.4\pm1.6$ \\
 $\gamma J/\psi$ & $0.8\pm0.4$ \\
 $\gamma \psi(2S)$ & $4.5\pm2.0$ \\
  \bottomrule[1pt]\bottomrule[1pt]
\end{tabular}
\end{table}

In order to pin down the nature of the $X(3872)$, searching for more
decay modes is crucial. In Table ~\ref{tab:br}, we list the observed
decays of the $X(3872)$. The dominant decay channel is the
open-charm decay, which is $37\%$ for the $ D^0 \bar{D}^{*0}$ and
$49\%$ for the $D^0 \bar{D}^{0}\pi^0$. The branching ratios of the
radiative decays $J/\psi\gamma$ and $\psi^\prime \gamma$ are of the
same order as those of the hidden-charm decays. Are there other
radiative decays of the $X(3872)$ whose decay rates could be as
large as those of the $J/\psi\gamma$ and $\psi^\prime \gamma$?

Recently, the LHCb Collaboration observed a sizeable $\omega$
contribution to $X(3872)\rightarrow J/\psi \pi\pi$
decay~\cite{LHCb:2022bly}. Inspired by the recent LHCb Collaboration
measurements, we study the $\rho$ and $\omega$ meson contributions
to the radiative decay processes $X(3872)\rightarrow J/\psi \pi
\gamma$ and $X(3872)\rightarrow J/\psi \pi\pi \gamma$ in this work.
In Ref.~\cite{Wang:2022vjm}, the authors noted that the dominant
contributions to $X(3872)\to J/\psi \pi^+ \pi^-$ and
$X(3872)\rightarrow J/\psi \pi^+ \pi^- \pi^0$ arise from the
diagrams with the $X(3872)$ coupling to the $J/\psi\rho$ and
$J/\psi\omega$, respectively. One may wonder whether the same
scenario still holds in the $X(3872)\rightarrow J/\psi \pi \gamma$
and $X(3872)\rightarrow J/\psi \pi \pi \gamma$.

Compared with $X(3872)\rightarrow J/\psi \pi \pi$,
$X(3872)\rightarrow J/\psi \pi \gamma$ has an advantage in exploring
the isospin violation of the $J/\psi\rho$ mode. The LHCb experiment
has proved that there is a sizeable $\omega$ contribution to
$X(3872)\rightarrow J/\psi \pi\pi$. In other words, the
$X(3872)\rightarrow J/\psi \pi\pi$ is not a clean process to study
the isospin violation of the $J/\psi\rho$ mode. In
Fig.~\ref{Fig:Tri1}(a), the $X(3872)\rightarrow J/\psi \pi \gamma$
decay occurs through the intermediate $\rho$ or $\omega$ meson. The
$\omega$ meson dominates this process because $g_{X\psi\omega}$ and
$g_{\omega\pi\gamma}$ are both much larger than $g_{X\psi\rho}$ and
$g_{\rho\pi\gamma}$ respectively. Thus, the $X(3872)\rightarrow
J/\psi \pi \gamma$ should be a cleaner process to extract the
coupling $g_{X\psi\omega}$. By the same token, the
$X(3872)\rightarrow J/\psi \pi \pi\gamma$ is a cleaner process to
study the isospin violation channel of $J/\psi\rho$. For this
purpose, we will not only check the contribution of the $\rho$ and
$\omega$ mesons to the $X(3872)\rightarrow J/\psi \pi \gamma$
process but also the contributions of diagrams with the $X(3872)$
coupling to the $J/\psi\rho$ or $J/\psi\omega$ in the
$X(3872)\rightarrow J/\psi \pi \pi\gamma$ process. Besides the
$\rho$ and $\omega$ contributions, there are some nonresonant
contributions which should be considered as the background
contribution. We will predict the branching ratios of
$X(3872)\rightarrow J/\psi \pi \gamma$ and $X(3872)\rightarrow
J/\psi \pi\pi \gamma$, which could be tested by the BESIII and LHCb
Collaborations.

This paper is organized as follows. After the introduction, we
present the theoretical framework in the calculation of
$X(3872)\rightarrow J/\psi \pi \gamma$ and $X(3872)\rightarrow
J/\psi \pi\pi \gamma$. We derive the invariant decay amplitudes and
invariant mass distributions using the effective Lagrangian method.
In Sec.~\ref{sec:results}, we present the invariant mass
distribution of $\pi \gamma$ and $\pi \pi \gamma$, and the branching
ratios of $X(3872)\rightarrow J/\psi \pi \gamma$ and
$X(3872)\rightarrow J/\psi \pi\pi \gamma$. Sec.~\ref{sec:summary} is
a short summary.

\section{Theoretical framework}
\label{sec:Sec2}

In this work, we utilize the effective Lagrangian method to study
the radiative processes $X(3872)\rightarrow J/\psi \pi \gamma$ and
$X(3872)\rightarrow J/\psi \pi\pi \gamma$. In the following
subsections, we introduce the effective Lagrangian and invariant
decay amplitudes and the formulas of the invariant mass
distributions related to the radiative processes $X(3872)\rightarrow
J/\psi \pi \gamma$ and $X(3872)\rightarrow J/\psi \pi\pi \gamma$.

\subsection{Feynman diagrams and effective Lagrangian}

\begin{figure}[htb]
\begin{tabular}{ccc}
  \centering
 \includegraphics[width=4.3cm]{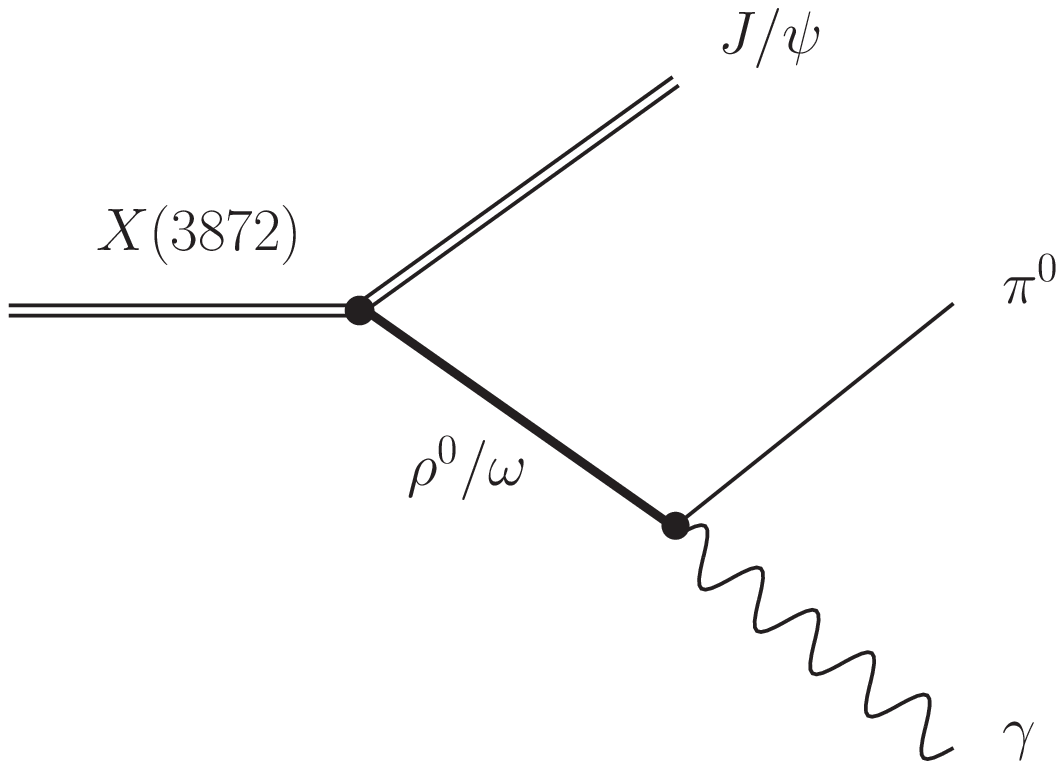}& \includegraphics[width=4.3cm]{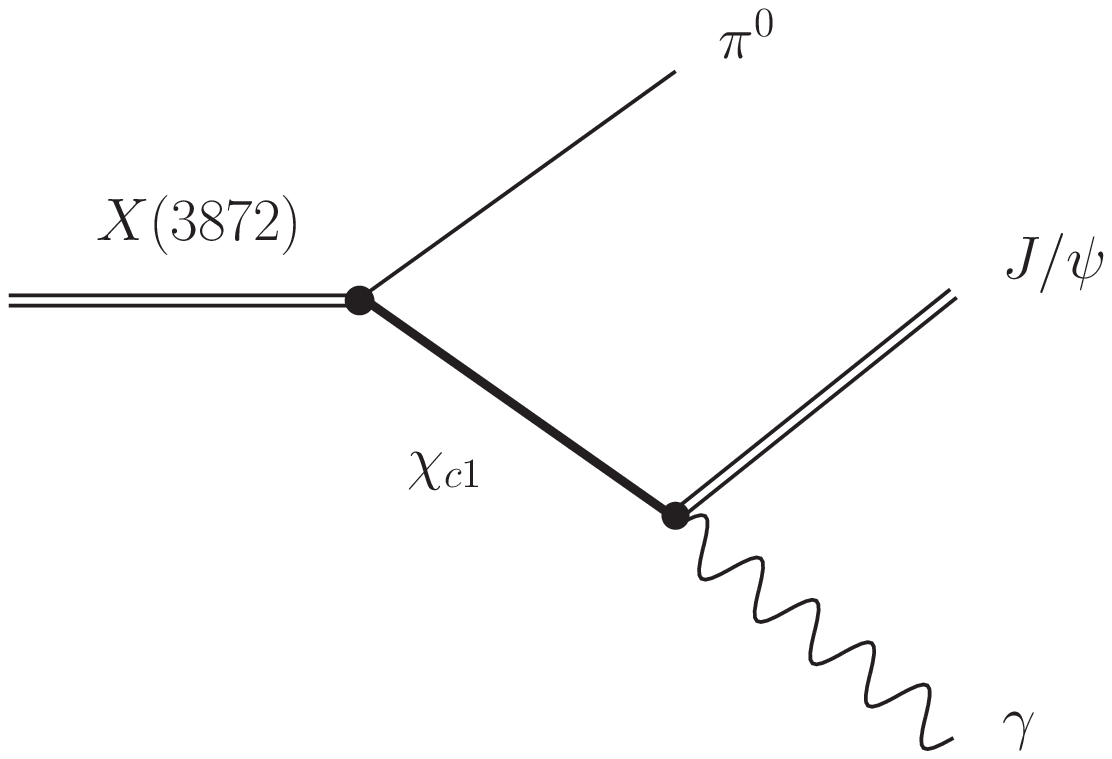} \\
 $(a)$ & $(b)$\\
 \end{tabular}
\caption{Diagrams for the $X(3872)\rightarrow J/\psi \pi \gamma$
with the $\rho^0/\omega$ contribution (a) and $\chi_{c1}$
contribution (b).}\label{Fig:Tri1}
\end{figure}

\begin{figure}[htb]
\begin{tabular}{ccc}
  \centering
 \includegraphics[width=4.4cm]{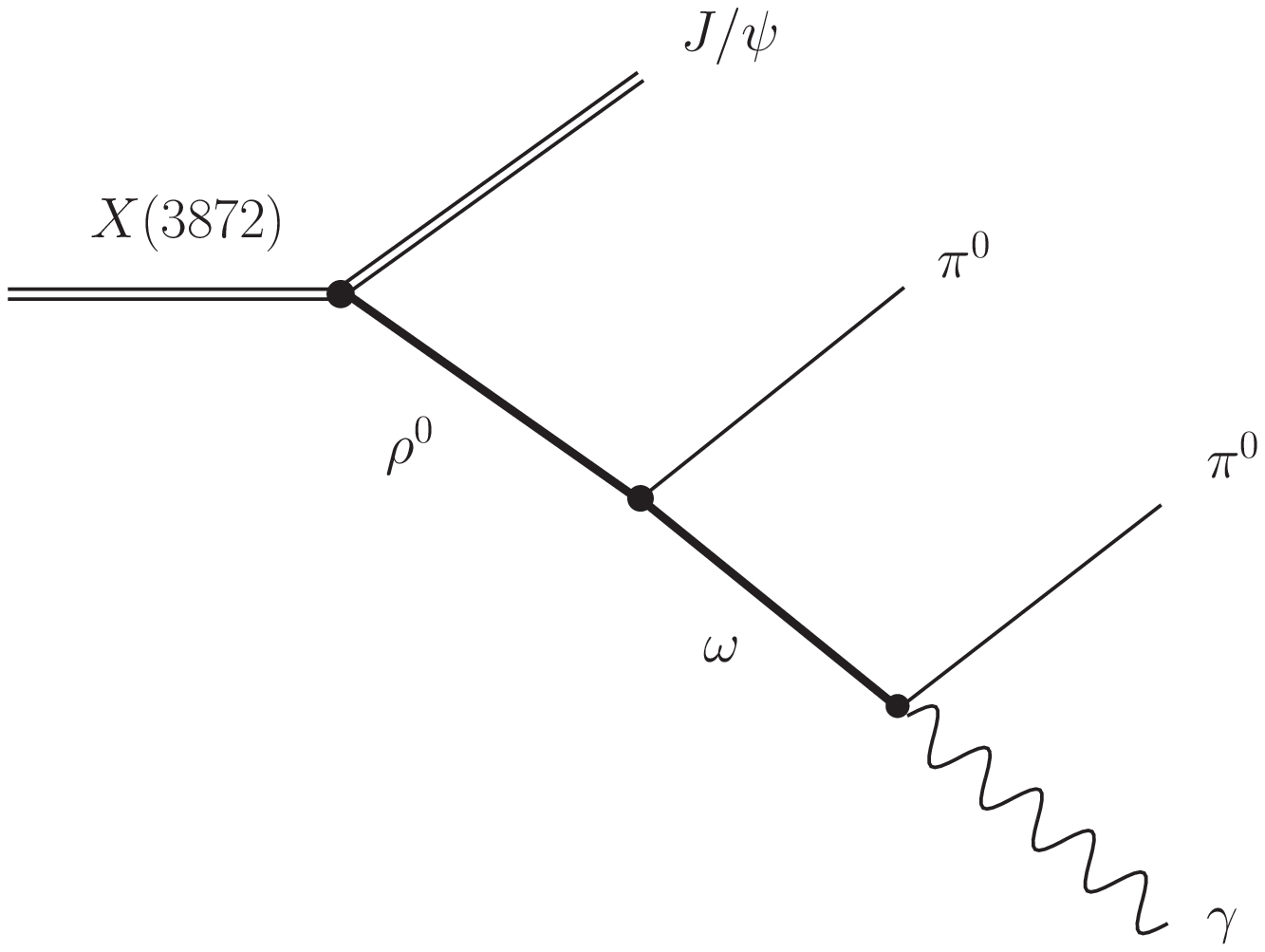}&  \includegraphics[width=4.4cm]{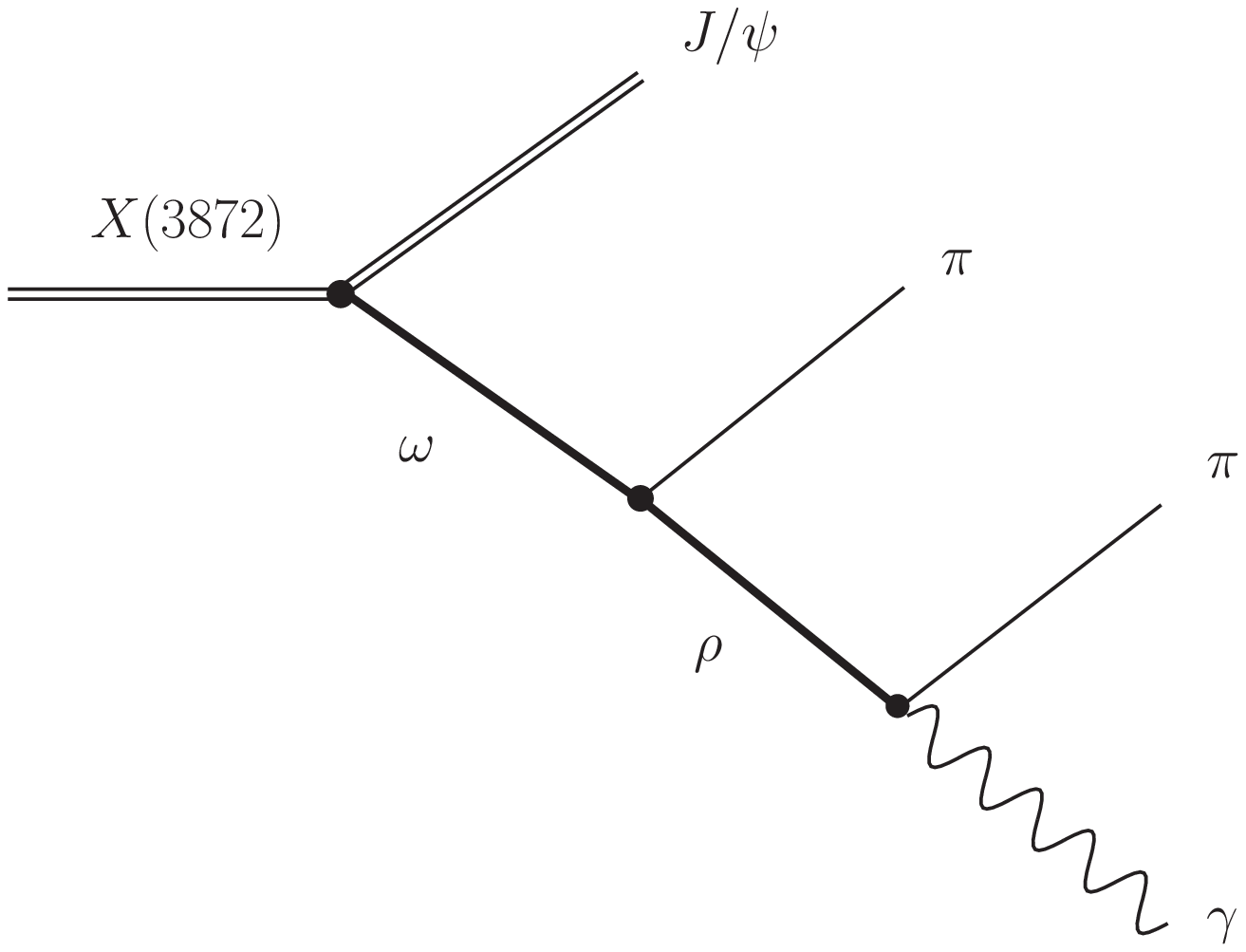}\\
 $(a)$ &  $(b)$ \\
  \includegraphics[width=4.4cm]{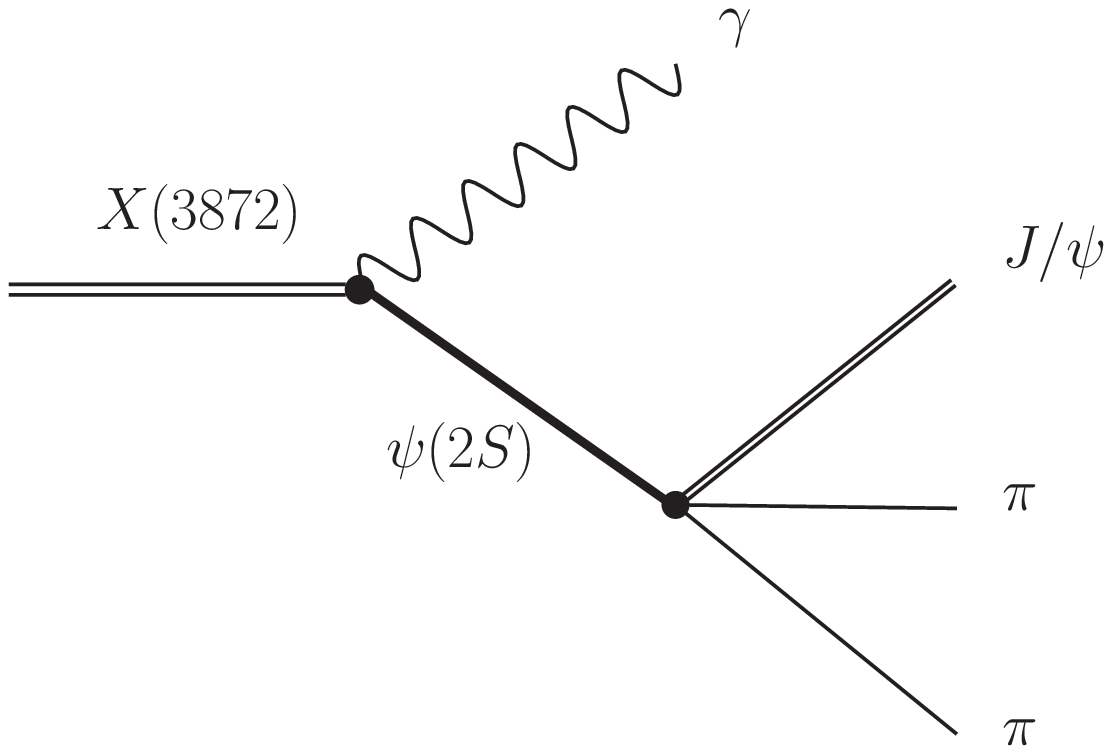}&  \includegraphics[width=4.4cm]{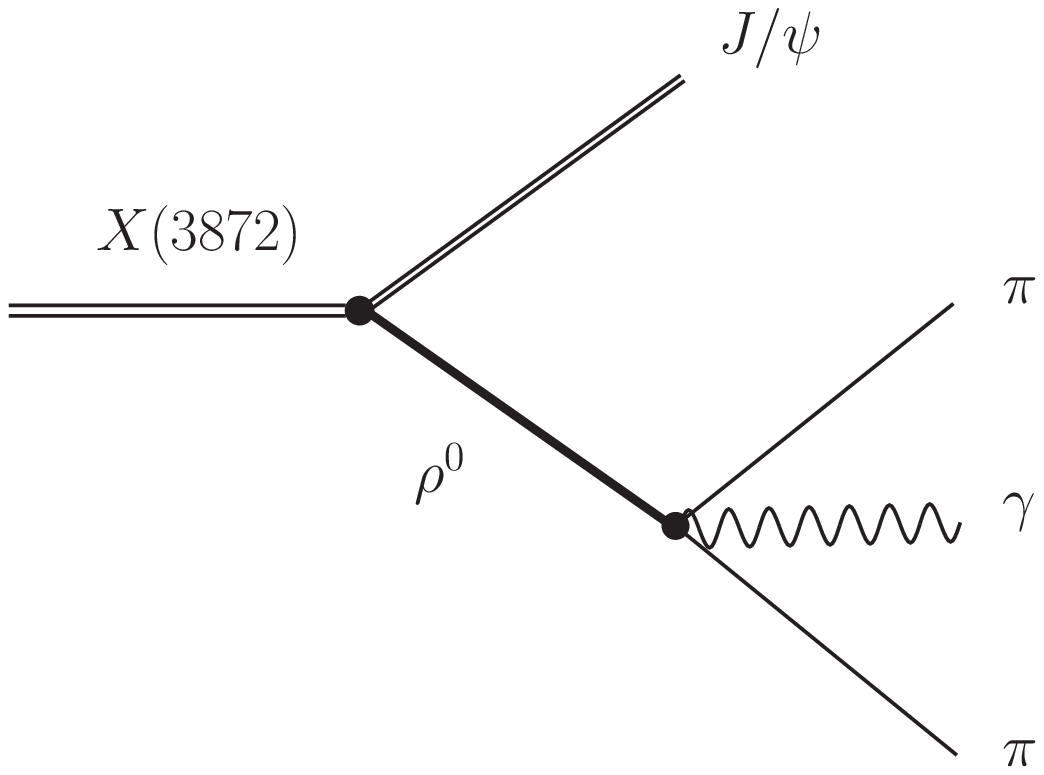}\\
  $(c)$ & $(d)$ \\
 \end{tabular}
\caption{Diagrams for the $X(3872)\rightarrow J/\psi \pi\pi \gamma$
with the $X(3872)$ coupling to the $J/\psi\rho$ (a), $J/\psi\omega$
(b), $\psi(2S)$ (c) and $\rho$ (d) respectively.}\label{Fig:Tri2}
\end{figure}

In Fig.~\ref{Fig:Tri1}(a) and Figs.~\ref{Fig:Tri2}(a)-(b), the
decays $X(3872)\rightarrow J/\psi \pi \gamma$ and
$X(3872)\rightarrow J/\psi \pi \pi\gamma$ occur through the $\rho$
and $\omega$ as the intermediate states. As shown in Table
~\ref{tab:br}, the branching ratios of the decays $X(3872)\to
\pi^0\chi_{c1}$ and $X(3872)\to \gamma\psi(2S)$ are sizable. In
addition, the branching ratios of $\psi(2S)\to J/\psi\pi^+ \pi^-$
and $\psi(2S)\to J/\psi\pi^0 \pi^0$ are $(34.68
\pm0.30)\%$ and $(18.24 \pm0.31)\%$ respectively~\cite{Workman:2022ynf}.
The branching ratio of the $\chi_{c1}\to \gamma J/\psi$ is also
quite large. Thus, the diagrams Fig.~\ref{Fig:Tri1}(b) and
Fig.~\ref{Fig:Tri2}(c) will also contribute to the background. In
contrast, the $\pi \gamma$ and $\pi \pi \gamma$ invariant mass
spectrum tend to peak around the $\rho$ and $\omega$ mass for our
concerned $\rho$ and $\omega$ contributions. Besides, the QED gauge
invariance requires the existence of Fig.~\ref{Fig:Tri2}(d).
One notes that the $\omega$ may also contribute to
Fig.~\ref{Fig:Tri2}(d). The branching ratio of
$\omega\rightarrow\pi^0\pi^0\gamma$ is $(6.7\pm 1.1)\times 10^{-5}$~\cite{Workman:2022ynf}.
The branching ratio of $\omega\rightarrow\pi^+\pi^-\gamma$ has not been
measured yet. If one neglects the long range contributions and
considers the isospin symmetry, the branching ratio
of $\omega\rightarrow\pi^+\pi^-\gamma$ is just twice the
$\omega\rightarrow\pi^0\pi^0\gamma$. In contrast, the branching
ratio of $\rho\rightarrow\pi^+\pi^-\gamma$ is around $10^{-2}$. In
other words, the $\omega$ contribution to Fig.~\ref{Fig:Tri2}(d) is
much smaller than the $\rho$ contribution. Thus, we only consider
the diagram in Fig.~\ref{Fig:Tri2}(d).

In order to get the invariant decay amplitudes in
Figs.~\ref{Fig:Tri1}-\ref{Fig:Tri2}, we need the following effective
Lagrangian~\cite{Janssen:1994uf,Lucio-Martinez:2000now,Casalbuoni:1992yd,DeFazio:2008xq},
\begin{eqnarray}
\mathcal{L}_{XJ/\psi V} &=& g_{X\psi V}\varepsilon^{\mu\nu\alpha\beta}\partial_{\mu}X_{\nu}\psi_{\alpha} V_{\beta}, \label{eq:Lag1} \\
\mathcal{L}_{X\chi_{c1} \pi} &=& \frac{g_{X\chi_{c1} \pi}}{m_X}\varepsilon^{\mu\nu\alpha\beta}\partial_{\mu}X_{\nu}\chi_{c1\alpha} \partial_\beta \pi, \label{eq:Lag2} \\
\mathcal{L}_{X\psi^\prime\gamma} &=& g_{X\psi^\prime\gamma} \varepsilon^{\mu\nu\alpha\beta} X_\mu \psi^\prime_\nu \partial_\alpha A^\gamma_\beta, \label{eq:Lag3} \\
\mathcal{L}_{\omega\rho\pi} &=& g_{\omega\rho\pi}\varepsilon^{\mu\nu\alpha\beta}\partial_{\mu}\omega_{\alpha}\partial_{\nu}\rho_{\beta}\phi_\pi,  \label{eq:Lag4} \\
\mathcal{L}_{V\pi\gamma} &=& g_{V\pi\gamma} \varepsilon^{\mu\nu\alpha\beta} F_{\mu\nu}V_{\alpha\beta}\phi_\pi, \label{eq:Lag5} \\
\mathcal{L}_{\chi_{c1}\psi\gamma} &=&g_{\chi_{c1}\psi\gamma}\varepsilon^{\mu\nu\alpha\beta}\partial_{\mu}\chi_{c1\nu}v^\xi \psi_{\alpha} F_{\beta\xi}, \label{eq:Lag6} \\
\nonumber
\end{eqnarray}
where $X$, $V$, $\psi^\prime$, $\chi_{c1}$ stand for $X(3872)$,
$\rho/\omega$, $\psi(2S)$ and $\chi_{c1}(1P)$, respectively.
$g_{X\psi V}$, $g_{X\chi{c1}\pi}$, $g_{X\psi^\prime\gamma}$,
$g_{\omega\rho\pi}$, $g_{V\pi\gamma}$ and $g_{\chi_{c1}\psi\gamma}$
are the relevant coupling constants and will be discussed in the
next subsection. In addition, the electromagnetic field strength
tensor is $F_{\mu\nu}=\partial_\mu A^\gamma_\nu-\partial_\nu
A^\gamma_\mu$ and $V_{\alpha\beta}=\partial_\alpha
V_\beta-\partial_\beta V_\alpha$.

The $\rho\to\pi\pi$ effective Lagrangian reads
\begin{eqnarray}
\mathcal{L}_{\rho\pi\pi}&=&g_{\rho\pi\pi}\rho_\mu
(\phi_{\pi^+}D^{\dagger\mu}\phi_{\pi^-}-\phi_{\pi^-}D^\mu\phi_{\pi^+}),
\end{eqnarray}
where $D^\mu =\partial^\mu+ieA^\mu$. The $\rho\to\pi\pi\gamma$
vertex arises from the contact seagull interaction, which
contributes to Fig. \ref{Fig:Tri2}(d).

\subsection{Invariant decay amplitudes}

With the above effective Lagrangian, the invariant decay amplitudes
of $X(p)\rightarrow J/\psi(p_1)+\rho(q)\rightarrow J/\psi(p_1)
+\pi(p_2) +\gamma(p_3)$ shown in Fig.~\ref{Fig:Tri1}(a) is
\begin{eqnarray}
\mathcal{M}^{\pi\gamma}_\rho&=&\Big(g_{X\psi\rho} \varepsilon_{\xi\kappa\phi\theta} i p^\xi \epsilon^\kappa(p)p^\phi_1 q^\theta\Big)\frac{-g^{\theta\sigma}+q^\theta q^\sigma/m^2_\rho}{D_\rho(q^2)}\nonumber\\
&&\times\Big(g_{\rho\pi\gamma}\varepsilon_{\mu\nu\alpha\beta}(p^\mu_3 g^{\rho\nu}-p^\nu_3 g^{\rho\mu})(q^\alpha g^{\sigma\beta}-q^\beta g^{\sigma\alpha})\epsilon_\rho(p_3)\Big)\nonumber\\
&&\times F_\rho(q^2),
\end{eqnarray}
the invariant decay amplitudes of $X(p)\rightarrow
\pi(p_2)+\chi_{c1}(q)\rightarrow J/\psi(p_1)+ \pi(p_2) +\gamma(p_3)$
shown in Fig.~\ref{Fig:Tri1}(b) is
\begin{eqnarray}
\mathcal{M}^{\pi\gamma}_{\chi_{c1}}&=&\Big(\frac{g_{X\chi_{c1} \pi}}{m_X} \varepsilon_{\xi\kappa\phi\theta} i p^\xi \epsilon^\kappa(p)i p^\theta_2\Big)\frac{-g^{\phi\nu}+q^\phi q^\nu/m^{\chi_{c1}}_\rho}{D_{\chi_{c1}}(q^2)}\nonumber\\
&&\times\Big(\frac{\sqrt{2}g_{\chi_{c1}\psi\gamma}}{m_{\chi_{c1}}}\varepsilon_{\mu\nu\alpha\beta}(-i)q^\mu v^\xi \epsilon^\alpha(p_1)i(p^\beta_3 \epsilon^\xi(p_3)\nonumber\\
&& -p^\xi_3 \epsilon^\beta(p_3))\Big)F_{\chi_{c1}}(q^2),
\end{eqnarray}
and the invariant decay amplitudes of $X(p)\rightarrow J/\psi(p_4)+
\pi(p_1)+\pi(p_2)+ \gamma(p_3)$ shown in Fig.~\ref{Fig:Tri2}(a) is
\begin{eqnarray}
\mathcal{M}^{\pi\pi\gamma}_\rho&=&\Big(g_{X\psi\rho} \varepsilon_{\xi\kappa\phi\theta} i p^\xi \epsilon^\kappa(p) \epsilon^\phi(p_4)\Big)\frac{-g^{\theta\eta}+q^\theta_1 q^\eta_1 /m^2_\rho}{D_\rho(q^2)}\nonumber\\
&&\times\Big(g_{\omega\rho\pi}\varepsilon_{\lambda\omega\delta\eta}q^\lambda_2 q^\omega_1\Big)\frac{-g^{\delta\sigma}+q^\delta_2 q^\sigma_2 /m^2_\omega}{D_\omega(q^2_2)}\nonumber\\
&&\times\Big(g_{\omega\pi\gamma}\varepsilon_{\mu\nu\alpha\beta}\epsilon_\rho(p_3)(p^\mu_3 g^{\rho\nu}-p^\nu_3 g^{\rho\mu})(q^\alpha_2 g^{\sigma\beta}-q^\beta_2 g^{\sigma\alpha})\Big)\nonumber\\
&&\times F_\rho(q^2_1)F_\omega(q^2_2),
\end{eqnarray}
where $p$, $p_1$, $p_2$, $p_3$, $p_4$ are the four-momenta of
$X(3872)$, $\pi^-$, $\pi^+$, $\gamma$, $J/\psi$, while $q_1$ and
$q_2$ represent the four-momenta of the intermediate $\rho^0$ and
$\omega$ mesons. $D_{\rho}(q^2)$ and $D_{\omega}(q^2_1)$ are the
denominators of the propagators for the $\rho$ and $\omega$ meson,
which are
\begin{eqnarray}
D_{\rho}(q^2_1) & =& q^2_1-m_{\rho}^2+im_{\rho}\Gamma_{\rho},  \label{Eq:rhowidth} \\
D_{\omega}(q^2_2) &=& q^2_2 -m^2_{\omega} + i m_{\omega}
\Gamma_{\omega}.
\end{eqnarray}
Here, the $\rho$ meson is not far away from its mass shell and the width of the
$\omega$ meson is narrow enough that its energy dependence can be safely neglected. Thus, we take the $\Gamma_{\rho/\omega}$ as a constant.

The invariant decay amplitudes of $X(3872)\rightarrow
J/\psi\omega\rightarrow J/\psi \pi \gamma$ in Fig.~\ref{Fig:Tri1}(a)
and $X(3872)\rightarrow J/\psi \pi \pi\gamma$ in
Fig.~\ref{Fig:Tri2}(b) can be obtained by
\begin{eqnarray}
\mathcal{M}^{\pi\gamma}_\omega=\mathcal{M}^{\pi\gamma}_\rho \mid_{g_{X\psi\rho}\to g_{X\psi\omega}, g_{\rho\pi\gamma}\to g_{\omega\pi\gamma}, m_\rho\to m_\omega},\nonumber\\
\mathcal{M}^{\pi\pi\gamma}_\omega=\mathcal{M}^{\pi\pi\gamma}_\rho
\mid_{g_{X\psi\rho}\to g_{X\psi\omega}, g_{\omega\pi\gamma}\to
g_{\rho\pi\gamma},m_\rho\leftrightarrow m_\omega}.
\end{eqnarray}

In evaluating the decay amplitudes of $X(3872) \to J/\psi \pi\gamma$
and $X(3872) \to J/\psi \pi\pi\gamma$ associated with the $\rho$ and
$\omega$ mesons in Figs.~\ref{Fig:Tri1} and \ref{Fig:Tri2}, we
include the form factors for the $\rho$ and $\omega$ mesons since
they are not point-like particles~\cite{Liu:1995st}. In this work we
adopt the following form factor:
\begin{eqnarray}
F_{\rho/\omega} (q^2) =
\frac{\Lambda_{\rho/\omega}^4}{\Lambda_{\rho/\omega}^4+(q^2-m_{\rho/\omega}^2)^2},
\label{eq:FF}
\end{eqnarray}
where we adopt $\Lambda_{\rho} = \Lambda_\omega = 598$ MeV as a
result of $\Gamma_\rho$ being a constant~\cite{Wang:2022vjm}. We
have checked that our results barely depend on the form factor.

In Ref.~\cite{Wang:2022vjm}, the coupling constants $g_{X\psi\rho}$
and $g_{X\psi\omega}$ are determined to be $0.09\pm0.02$ and
$0.31\pm0.06$ by fitting to the LHCb data with $\Gamma_\rho$ being a
constant. Other coupling constants can be determined from the
corresponding experimental partial widths. With the effective
Lagrangian in Eqs.~(\ref{eq:Lag2})-(\ref{eq:Lag6}), the decay widths
of $\rho\to \pi\gamma$, $\omega\to \pi\gamma$, $X(3872) \to
\psi(2S)\gamma$, $X(3872) \to \chi_{c1}\pi$ and $\chi_{c1}\to
J/\psi\gamma$ are
\begin{eqnarray}
\Gamma_{\rho\to \pi\gamma}&=&\frac{4g^2_{\rho\pi\gamma}p^3_{f\rho}}{3\pi},\\
\Gamma_{\omega\to \pi\gamma}&=&\frac{4g^2_{\omega\pi\gamma}p^3_{f\omega}}{3\pi},\\
\Gamma_{X\to \psi^\prime\gamma}&=&\frac{g^2_{X\psi^\prime\gamma}p^3_{fX}}{12\pi m^2_X m^2_{\psi^\prime}}\Big(m^2_X + m^2_{\psi^\prime}\Big),\\
\Gamma_{X\to \chi_{c1}\pi}&=&\frac{g^2_{X\psi^\prime\gamma}p^3_{fX}}{12\pi m^2_X},\\
\Gamma_{\chi_{c1}\to
J/\psi\gamma}&=&\frac{g^2_{\chi_{c1}\psi\gamma}p^3_{f\chi_{c1}}m_\psi}{3\pi
m_{\chi_{c1}}},\\ \nonumber
\end{eqnarray}
where $p_{f\rho}$, $p_{f\omega}$, $p_{fX}$ and $p_{f\chi_{c1}}$ are
the three momenta of the final mesons in the $\rho$, $\omega$,
$X(3872)$ and $\chi_{c1}$ rest frame, respectively. With
$\mathcal{B}[\rho^0\to \pi^0\gamma]=4.7\times10^{-4}$,
$\mathcal{B}[\omega\to \pi^0\gamma]=8.35\%$, $\mathcal{B}[X(3872)\to
\psi^\prime\gamma]=4.5\%$ and $\mathcal{B}[X(3872)\to
\chi_{c1}\pi]=3.4\%$, we have $|g_{\rho\pi\gamma}|=0.06
\mathrm{GeV}^{-1}$, $|g_{\omega\pi\gamma}|=0.18 \mathrm{GeV}^{-1}$,
$|g_{X\psi^\prime\gamma}|=1.56$ and
$|g_{X\chi{c1}\pi}|=0.84^{+0.18}_{-0.23}$.
$g_{\chi_{c1}\psi\gamma}=\sqrt{\frac{2m_\psi}{m_{\chi_{c1}}}}g_{PS\gamma}$
and $|g_{PS\gamma}|=0.23 \mathrm{GeV}^{-1}$. $g_{\omega\rho\pi}$ can
be determined from the experimentally measured partial decay of
$\omega\to \rho\pi \to \pi\pi\pi$, which is
$|g_{\omega\rho\pi}|=50\mathrm{GeV}^{-1}$ with $\Gamma_\rho$ being a
constant~\cite{Wang:2022vjm}. Note that one can only obtain the
absolute value of the coupling constant from the partial decay
width. The phase can not be fixed. In this work, the default values
of the above coupling constants are real and positive.

The total invariant decay amplitudes of $X(3872)\rightarrow J/\psi
\pi \gamma$ and $X(3872)\rightarrow J/\psi \pi\pi \gamma$ are
\begin{eqnarray}
\mathcal{M}_{X\rightarrow J/\psi \pi \gamma}&=&\mathcal{M}^{\pi\gamma}_\rho+ e^{i\phi_{\omega}}\mathcal{M}^{\pi\gamma}_\omega + e^{i\phi_{\chi_{c1}}}\mathcal{M}^{\pi\gamma}_{\chi_{c1}}, \notag\\
\mathcal{M}_{X\rightarrow J/\psi \pi\pi
\gamma}&=&\mathcal{M}^{\pi\pi\gamma}_\rho+
e^{i\phi_\omega}\mathcal{M}^{\pi\pi\gamma}_\omega
+
e^{i\phi_{\psi^\prime}}\mathcal{M}^{\pi\pi\gamma}_{\psi^\prime},
\end{eqnarray}
where $\phi_\omega$ stands for the relative phase between the
$\omega$ and $\rho$ terms, $\phi_{\psi^\prime}$ stands for the
relative phase between $\rho/\omega$ and $\psi^\prime$ terms. We
adopt the phase angle $\phi_\omega$ obtained by fitting the LHCb data in
Ref.~\cite{Wang:2022vjm}, which is $134.5^\circ$.

\subsection{Invariant mass distributions}

The invariant $\pi^0\gamma$ mass distribution of the $X(3872) \to
J/\psi \pi^0\gamma$ decay is given by
\begin{eqnarray}
\frac{\text{d}\Gamma_{X{3872} \to J/\psi \pi^0 \gamma}}{\text{d}M_{\pi^0 \gamma}} &=& \frac{1}{24(2\pi)^4M_X^2} \notag\\
&&\times\int\Sigma|\mathcal{M}_{\pi\gamma}|^2|\bold{p}_1^{\ast}||\bold{p}_4|\text{d}\cos{\theta_1}\text{d}\phi_1,
\label{formula:dgdm23}
\end{eqnarray}
where $\bold{p}_1^{\ast}$ and ($\theta_1$, $\phi_1$) are the
three-momentum and decay angle of the outgoing $\pi^0/\gamma$ in the
center-of-mass (c.m.) frame of the final $\pi^0 \gamma$ system,
$\bold{p}_4$ is the three-momentum of the final $J/\psi$ meson in
the rest frame of $X(3872)$, and $M_{\pi^0 \gamma}$ is the invariant
mass of the final $\pi^0 \gamma$ system.

For the invariant $\pi\pi\gamma$ mass distributions of the $X(3872)
\to J/\psi \pi\pi\gamma$ decay,
    \begin{eqnarray}
     &&\frac{\text{d}\Gamma_{X(3872) \to J/\psi \pi\pi\gamma}}{\text{d}M_{\pi\pi\gamma}} = \frac{1}{16(2\pi)^7M_X^2} \int\Sigma|\mathcal{M}_{\pi\pi\gamma}|^2  \times \notag \\
     && |\bold{p}_1^{\ast}||\bold{p}_3'||\bold{p}_4| \text{d}M_{\pi\gamma}\text{d}\cos{\theta_1}\text{d}\phi_1\text{d}\cos{\theta_2}\text{d}\phi_2,  \label{formula:dgdm123}
\end{eqnarray}
with $M_{\pi\pi\gamma}$ the invariant mass of $\pi\pi\gamma$ system.
The $\bold{p}_1^{\ast}$ and ($\theta_1$, $\phi_1$) are the
three-momentum and decay angles of the outgoing $\pi$ in the
$\pi\gamma$ center-of-mass (c.m.) frame. The $\bold{p}'_3$ and
($\theta_2$, $\phi_2$) are the three-momentum and decay angles of
the outing $\pi^0$ in the $\pi\pi\gamma$ c.m. frame. The
$\bold{p}_4$ is the three-momentum of the final $J/\psi$ meson in
the $X(3872)$ rest frame. Definitions of these variables in the
phase space integration of the $X(3872) \to J/\psi \pi\pi\gamma$
decay can be found in the Appendix of Ref.~\cite{Wang:2022vjm}.

\section{Numerical results and discussion}
\label{sec:results}

\subsection{$X(3872)\rightarrow J/\psi \pi \gamma$}

\begin{figure}[htb]
  \centering
 \includegraphics[width=7.5 cm]{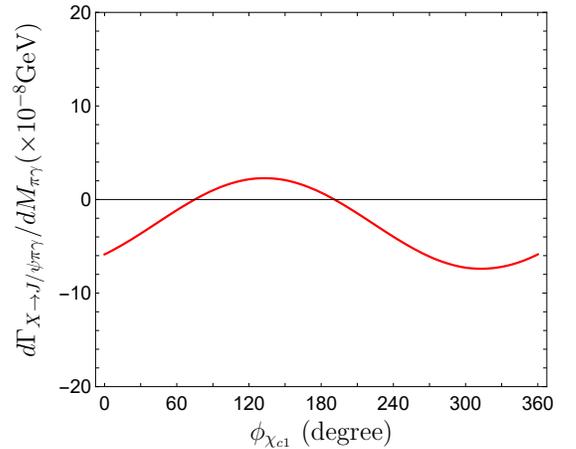}
  \caption{The $\phi_{\chi_{c1}}$ dependence of the invariant mass distribution of $\pi \gamma$ for the $X(3872)\rightarrow J/\psi \pi \gamma$ by considering the interference term with $m_{\pi\gamma}=0.5$ GeV.}\label{Fig:imd0}
\end{figure}

In this work, we assume that the phase angle $\phi_\omega$ in
$X(3872)\rightarrow J/\psi \pi \gamma$ is the same as in the
$X(3872)\rightarrow J/\psi \pi \pi$. Unfortunately, due to the
absence of the experimental data, the other phase $\phi_{\chi_{c1}}$
is unknown. We first investigate the $\phi_{\chi_{c1}}$ dependence
of the interference term by setting the $m_{\pi\gamma}=0.5$ GeV,
which is shown in Fig.~\ref{Fig:imd0}. One can see that the
interference term is not drastically dependent on the
$\phi_{\chi_{c1}}$. Thus it is reasonable to choose the phase angle
$\phi_{\chi_{c1}}=223^\circ$ to estimate the invariant mass
distribution of $\pi \gamma$ for the $X(3872)\rightarrow J/\psi \pi
\gamma$, which corresponds to the central value of the interference
term.

\begin{figure}[htb]
  \centering
 \includegraphics[width=7.5 cm]{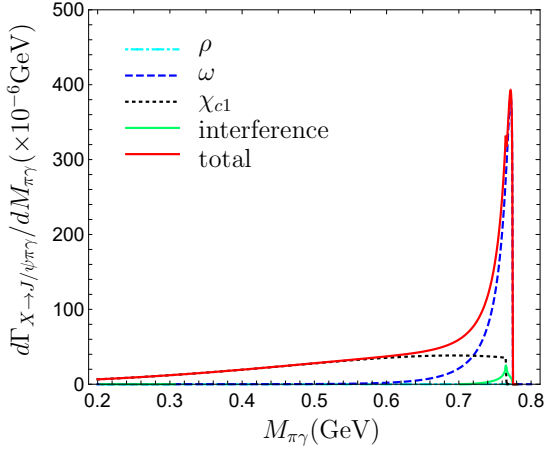}
  \caption{Invariant mass distribution of $\pi \gamma$ for the $X(3872)\rightarrow J/\psi \pi \gamma$. The blue-dash-dotted, blue-dashed, black-dashed, green and red solid are the $\rho$, $\omega$, $\chi_{c1}$, interference term and total contribution, respectively.}\label{Fig:imd1}
\end{figure}

In Fig.~\ref{Fig:imd1}, we present the invariant mass distribution
of $\pi \gamma$ for the $X(3872)\rightarrow J/\psi \pi \gamma$ when
the $\phi_\omega$ and $\phi_{\chi_{c1}}$ are both fixed. Different
from the $X(3872)\rightarrow J/\psi \pi \pi$ in
Ref.~\cite{Wang:2022vjm} which is dominated by the $\rho$ meson, the
decay of $X(3872)\rightarrow J/\psi \pi \gamma$ is dominated by the
$\omega$ meson. The line shape of the $\omega$ contribution and the
total contribution are almost coincident in the high invariant mass
region. The differential decay rate with respect to $\pi \gamma$
from the $\omega$ contribution is two orders of magnitude larger
than that from the $\rho$ meson since $g_{X\psi\omega}$ and
$g_{\omega\pi\gamma}$ are both three times larger than
$g_{X\psi\rho}$ and $g_{\rho\pi\gamma}$ respectively. Thus, the
dominant resonance contribution of $X(3872)\rightarrow J/\psi \pi
\gamma$ is the $\omega$ meson. The $\chi_{c1}$ term provides the
dominant the non-resonance contribution, which serves as the
background. Due to the absolute dominance of the $\omega$ in
$X(3872)\rightarrow J/\psi \pi \gamma$, $X(3872)\rightarrow J/\psi
\pi \gamma$ becomes a clean and ideal process to explore the isospin
conservation channel $J/\psi\omega$ of $X(3872)$. In the line shape
of the total invariant mass distribution, there is a dip around
$766$ MeV, which results from the dip of the interference term.
After integrating over the $\pi \gamma$ invariant mass, the
branching ratio of $X(3872)\rightarrow J/\psi \pi \gamma$ is
$(8.10^{+3.44}_{-2.84})\times10^{-3}$ considering the $\rho$ and
$\omega$ contributions only.

\begin{figure}[htb]
  \centering
 \includegraphics[width=7.5 cm]{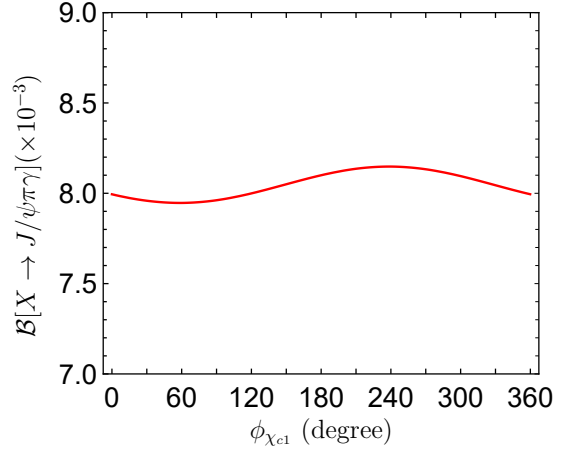}
  \caption{The $\phi_{\chi_{c1}}$ dependence of the total branching ratio of $X(3872)\rightarrow J/\psi \pi \gamma$.}\label{Fig:tot}
\end{figure}

The above branching ratio does not include the contribution from the
$\chi_{c1}$ term. To gain the total branching ratio of
$X(3872)\rightarrow J/\psi \pi \gamma$ including the $\chi_{c1}$
term, the $\phi_{\chi_{c1}}$ dependence of the total branching ratio
of $X(3872)\rightarrow J/\psi \pi \gamma$ should be clarified.

In Fig.~\ref{Fig:tot}, we present the $\phi_{\chi_{c1}}$ dependence
of the total branching ratio of $X(3872)\rightarrow J/\psi \pi
\gamma$ by fixing the $\phi_{\omega}$ to be $134.5^\circ $ and
varying the $\phi_{\chi_{c1}}$ from $0^\circ$ to $360^\circ$. The
$\phi_{\chi_{c1}}$ dependence of the total branching ratio of
$X(3872)\rightarrow J/\psi \pi \gamma$ is fairly stable. Finally,
the predicted branching ratio of $X(3872)\rightarrow J/\psi \pi
\gamma$ is $(8.10^{+3.59}_{-2.89})\times10^{-3}$. The central value
is obtained by taking $\phi_{\chi_{c1}}=180^\circ$, the errors come
from the variation of the $\phi_{\chi_{c1}}$. Under the assumption
that $X(3872)$ is a $D\bar{D}^*$ molecule and that its decay
proceeds through the transitions to $J/\psi\rho$ and $J/\psi\omega$,
the branching ratio of $X(3872)\rightarrow J/\psi \pi \gamma$ was
estimated to be $0.17\times\mathcal{B}[X\to
J/\psi\pi\pi]$~\cite{Braaten:2005ai}, which is similar to our
estimation. Our results indicate that the branching ratio of
$X(3872)\rightarrow J/\psi \pi \gamma$ is almost of the same order
as those of the hidden-charm and radiative decays to
$\psi^\prime/J/\psi$ of the $X(3872)$, which is large enough to be
detected experimentally.

\subsection{$X(3872)\rightarrow J/\psi \pi \pi\gamma$}

\begin{figure}[htb]
  \centering
 \includegraphics[width=7.5 cm]{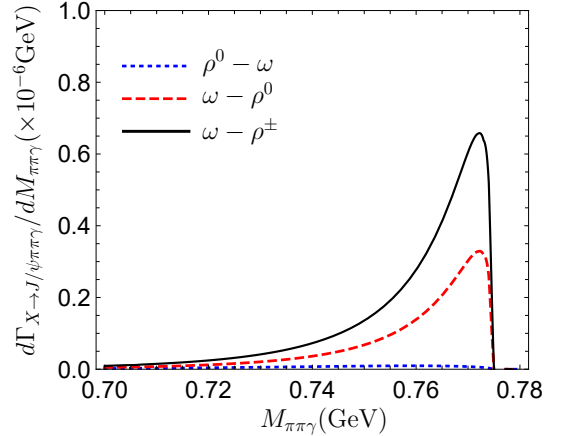}
\caption{Invariant mass distribution of $\pi\pi \gamma$ for the
$X(3872)\rightarrow J/\psi \pi \pi\gamma$. The blue-dotted stands
for $X(3872)\rightarrow J/\psi \pi^0 \pi^0\gamma$ with the
intermediate states $\rho^0-\omega$. The red-dashed and black-solid
stand for $X(3872)\rightarrow J/\psi \pi^0 \pi^0\gamma$ and
$X(3872)\rightarrow J/\psi \pi^+ \pi^-\gamma$ with the intermediate
states $\omega-\rho^0$ and $\omega-\rho^\pm$,
respectively.}\label{Fig:imd2}
\end{figure}

In the hidden charm decay of $X(3872)\rightarrow J/\psi \pi\pi\pi$,
the coupling constants $g_{X\psi\omega}$ and $g_{\rho\pi\pi}$ are
both larger than $g_{X\psi\rho}$ and $g_{\omega\pi\pi}$
respectively. As a result, the diagram where the $X(3872)$ couples
to $J/\psi\omega$ is far more important than the diagram where the
$X(3872)$ couples to the $J/\psi\rho$~\cite{Wang:2022vjm}.

For the radiative decay of $X(3872)\rightarrow J/\psi \pi\pi\gamma$,
$g_{X\psi\omega}$ is larger than $g_{X\psi\rho}$, while
$g_{\rho\pi\gamma}$ is smaller than $g_{\omega\pi\gamma}$ as shown
in Figs.~\ref{Fig:Tri2}(a)-(b). Thus, the contribution of Fig.
\ref{Fig:Tri2}(a) is probably comparable to that of Fig.
\ref{Fig:Tri2}(b). Here, it should be noted that Fig.
\ref{Fig:Tri2}(a) only contributes to the $X(3872)\rightarrow J/\psi
\pi^0\pi^0\gamma$ process. In Fig.~\ref{Fig:imd2}, we show the
results of the $\pi\pi\gamma$ invariant mass spectrum based on the
contributions of Fig.~\ref{Fig:Tri2}(a) and \ref{Fig:Tri2}(b), which are governed
by the $J/\psi\rho$ and $J/\psi\omega$ coupling, respectively. It
can be seen that the contribution of the $J/\psi\omega$ channel is
still larger than that of the $J/\psi\rho$ channel. After
integrating over the $\pi \pi \gamma$ invariant mass, the branching
ratios of $X(3872)\rightarrow J/\psi \pi^0\pi^0 \gamma$ are
$(3.84^{+1.90}_{-1.52})\times10^{-7}$ for Figs.~\ref{Fig:Tri2}(a)
and $(4.58^{+1.94}_{-1.60})\times10^{-6}$ for
Figs.~\ref{Fig:Tri2}(b).

In addition to Figs.~\ref{Fig:Tri2}(a)-(b), the diagram in
Fig.~\ref{Fig:Tri2}(c) could also contribute to $X(3872)\rightarrow
J/\psi \pi \pi\gamma$. The intermediate state $\psi(2S)$ is so
narrow that we can use the narrow width approximation to estimate
its contribution, which is $\mathcal{B}[X(3872) \to \gamma\psi(2S)
\to \gamma J/\psi\pi\pi]\simeq \mathcal{B}[X(3872) \to
\gamma\psi(2S)]\times \mathcal{B}[\psi(2S) \to J/\psi\pi\pi]$. Using
$\mathcal{B}[X(3872) \to \gamma\psi(2S)]=(4.5\pm2.0)\%$ and
$\mathcal{B}[\psi^\prime\to J/\psi\pi^0\pi^0]=(18.24 \pm0.31)\%$
given by PDG~\cite{Workman:2022ynf}, the branching ratios of
$X(3872)\rightarrow \gamma\psi^\prime\to \gamma J/\psi \pi^0\pi^0$
is $(0.82\pm0.37)\%$.

As for the direct coupling diagram in Fig.~\ref{Fig:Tri2}(d), the
intermediate $\rho$ meson is almost on shell with a large width.
Since the threshold of $J/\psi\rho$ is very close to the mass of
$X(3872)$, it is a good approximation to write the decay width of
$X(3872) \to \rho J/\psi \to J/\psi\pi\pi\gamma$ as
\begin{eqnarray}
\Gamma_{X\to  J/\psi \rho \to J/\psi\pi\pi\gamma}&=&\int^{(m_X-m_{J/\psi})^2}_{(2m_\pi)^2} ds f(s,m_\rho,\Gamma_\rho)\nonumber\\
&&\times\frac{|\vec{p}|}{24\pi m^2_X}\mid\overline{\mathcal{M}^{\mathrm{tot}}_{X\to  J/\psi \rho}(m_\rho\to \sqrt{s})}\mid^2 \nonumber\\
&&\times \mathcal{B}[\rho\to \pi\pi\gamma],\label{Eq:Gamma1}
\end{eqnarray}
which is equivalent to the appendix of Ref.~\cite{Meng:2021kmi}.
$f(s,m_\rho,\Gamma_\rho)$ is a
relativistic form of the Breit-Wigner distribution, which reads
\begin{eqnarray}
f(s,m_\rho,\Gamma_\rho)=\frac{1}{\pi}\frac{m_\rho
\Gamma_\rho}{(s-m^2_\rho)^2+m^2_\rho \Gamma^2_\rho},
\end{eqnarray}
and the amplitude $\mathcal{M}^{\mathrm{tot}}_{X(3872)\to  J/\psi
\rho}(m_\rho\to \sqrt{s})$ can be obtained by replacing the $\rho$
meson mass by $\sqrt{s}$. In the same way the momentum of the final state becomes,
\begin{eqnarray}
|\vec{p}|=\frac{\sqrt{[m^2_X-(\sqrt{s}-m_{J/\psi})^2][m^2_X-(\sqrt{s}+m_{J/\psi})^2]}}{2m_X}.
\end{eqnarray}

The invariant mass distribution of $\pi\pi\gamma$ for the $X(3872)
\to \rho J/\psi \to J/\psi\pi\pi\gamma$ decay is
\begin{eqnarray}
\frac{d\Gamma_{X\to  J/\psi \rho \to J/\psi\pi\pi\gamma}}{ds}&=&f(s,m_\rho,\Gamma_\rho)\nonumber\\
&&\times\frac{|\vec{p}|}{24\pi m^2_X}\mid\overline{\mathcal{M}^{\mathrm{tot}}_{X\to  J/\psi \rho}(m_\rho\to \sqrt{s})}\mid^2 \nonumber\\
&&\times \mathcal{B}[\rho\to \pi\pi\gamma].\label{Eq:Gamma2}
\end{eqnarray}
The branching ratios of $\rho\to \pi^+ \pi^-\gamma$ and $\rho\to
\pi^0 \pi^0\gamma$ are $(9.9\pm1.6)\times10^{-3}$ and
$(4.5\pm0.8)\times10^{-5}$~\cite{Workman:2022ynf}, respectively. In
this way, the branching ratio of $X(3872)\rightarrow J/\psi\rho\to
J/\psi \pi^0\pi^0\gamma$ is $(2.07\pm0.52)\times10^{-6}$.

Now we discuss the $X(3872)\rightarrow J/\psi \pi^+ \pi^- \gamma$.
After integrating over the $\pi \pi \gamma$ invariant mass, the
branching ratio of $X(3872)\rightarrow J/\psi \omega \rightarrow
J/\psi \pi^+\pi^- \gamma$ are $(9.16^{+3.89}_{-3.20})\times10^{-6}$
for Fig.~\ref{Fig:Tri2}(b). Using the narrow width approximation and
$\mathcal{B}[\psi^\prime\to J/\psi\pi^+\pi^-]=(34.68
\pm0.30)\%$~\cite{Workman:2022ynf}, the contribution of
Fig.~\ref{Fig:Tri2}(c) is $\mathcal{B}[X(3872) \to \gamma\psi(2S)
\to \gamma J/\psi\pi^+\pi^-]\simeq \mathcal{B}[X(3872) \to
\gamma\psi(2S)]\times \mathcal{B}[\psi(2S) \to
J/\psi\pi^+\pi^-]=(1.56\pm0.69)\%$.

With Eqs.~(\ref{Eq:Gamma1})-(\ref{Eq:Gamma2}), the branching ratio
of $X(3872)\rightarrow J/\psi\rho\to J/\psi \pi^+\pi^-\gamma$ is
estimated to be $(4.55\pm1.09)\times10^{-4}$ for
Fig.~\ref{Fig:Tri2}(d). In addition to the important background
contribution of $X(3872)\rightarrow \gamma\psi^\prime\to \gamma
J/\psi \pi\pi$, the $J/\psi \rho$ channel contribution is far larger
than that of the $J/\psi \omega$ channel in the $X(3872)\rightarrow
J/\psi \pi^+ \pi^- \gamma$. In other words, the radiative transition
of $X(3872)\rightarrow J/\psi \pi^+ \pi^- \gamma$ is a very clean
process to precisely study the isospin violation property of
$X(3872)$.

In the present estimation, all the involved coupling constants are
extracted from the corresponding experimental data. Thus, one should
get the same results regardless of the molecular or other scenarios
for the $X(3872)$. On the other hand, the $X(3872)\rightarrow J/\psi
\pi \gamma$ and $X(3872)\rightarrow J/\psi \pi \pi\gamma$ decays are
very helpful for constraining the coupling constants $g_{X\psi\rho}$
and $g_{X\psi\omega}$,
\begin{eqnarray}
&&\mathcal{B}[X\rightarrow J/\psi\pi\gamma]\nonumber\\
&=&0.002 g^2_{X\psi\rho}+0.083 g^2_{X\psi\omega}+0.004 g_{X\psi\rho}g_{X\psi\omega}+ 0.012,\nonumber\\
&&\mathcal{B}[X\rightarrow J/\psi\pi^0\pi^0\gamma]\nonumber\\
&=&(3.03\pm1.16)\times10^{-4} g^2_{X\psi\rho}+(4.77\times10^{-5}) g^2_{X\psi\omega}\nonumber\\
&&+(0.82\pm0.37)\%,\nonumber\\
&&\mathcal{B}[X\rightarrow J/\psi\pi^+\pi^-\gamma]\nonumber\\
&=&(5.62\pm2.22)\times10^{-2} g^2_{X\psi\rho}+(9.53\times10^{-5}) g^2_{X\psi\omega}\nonumber\\
&&+(1.56\pm0.69)\%.\nonumber\\
\end{eqnarray}

Note that we have assumed that the interference of the diagrams in
Fig.~\ref{Fig:Tri2} is negligible. From $\mathcal{B}[X\rightarrow
J/\psi\pi\gamma]$, the coefficient of $g^2_{X\psi\omega}$ is so
large that we can easily extract the coupling of $X\psi\omega$ in
$X\rightarrow J/\psi\omega \to J/\psi\pi\gamma$. The coefficients of
$g^2_{X\psi\rho}$ and $g^2_{X\psi\omega}$ in
$\mathcal{B}[X\rightarrow J/\psi\pi^0\pi^0\gamma]$ are pretty small
and thus it is difficult to obtain any useful information about
these couplings in $X\rightarrow J/\psi\pi^0\pi^0\gamma$. In
contrast, it is very interesting to see that the coefficient of
$g^2_{X\psi\rho}$ in $\mathcal{B}[X\rightarrow
J/\psi\pi^+\pi^-\gamma]$ is very large. Thus $X\rightarrow
J/\psi\pi^+\pi^-\gamma$ is a very good process to extract the
coupling $X\psi\rho$. We look forward to the measurement of the
branching ratios of $X(3872)\rightarrow J/\psi \pi \gamma$ and
$X(3872)\rightarrow J/\psi \pi\pi \gamma$ in the near future. At
that time, not only the predicted branching ratios can be tested,
but also the coupling constants $g_{X\psi\rho}$ and
$g_{X\psi\omega}$ can be extracted.

\section{Summary}
\label{sec:summary}

As the first established charmonium-like state, $X(3872)$ is one of
the best studied exotic hadron states both experimentally and
theoretically. Since its discovery, the mass spectrum, decay
behaviors and production mechanism of the $X(3872)$ have been
studied extensively. The $D\bar{D}^*$ hadronic molecule is the most
popular explanation, with which most of the phenomena related to
$X(3872)$ could be best explained. However, the other
interpretations can not be easily rule out.

In this work, we have studied the $\rho$ and $\omega$ meson
contribution to the radiative decays $X(3872)\rightarrow J/\psi \pi
\gamma$ and $X(3872)\rightarrow J/\psi \pi\pi \gamma$ using an
effective Lagrangian method. We obtain the invariant decay
amplitudes of the possible diagrams which contribute to
$X(3872)\rightarrow J/\psi \pi \gamma$ and $X(3872)\rightarrow
J/\psi \pi\pi \gamma$. We first investigate the $\phi_{\chi_{c1}}$
dependence of the interference term in $X(3872)\rightarrow J/\psi
\pi \gamma$, which is not drastic. Thus, we choose a central value
of $\phi_{\chi_{c1}}$ to analyse the invariant mass distribution of
$\pi \gamma$ for the $X(3872)\rightarrow J/\psi \pi \gamma$. The
total branching ratio of $X(3872)\rightarrow J/\psi \pi \gamma$
reaches $(8.10^{+3.59}_{-2.89})\times10^{-3}$, which barely depends
on $\phi_{\chi_{c1}}$.

Although the $\rho$ meson contribution is dominant in
$X(3872)\rightarrow J/\psi \pi \pi$, the $\omega$ contribution is
also sizable as recently measured by the LHCb Collaboration
\cite{LHCb:2022bly}. Our numerical results strongly indicate that
the $X(3872)\rightarrow J/\psi \pi \gamma$ is dominated by the
$\omega$ meson. Compared with $X(3872)\rightarrow J/\psi \pi \pi$,
$X(3872)\rightarrow J/\psi \pi \gamma$ is an ideal place to extract
the coupling of $X(3872)$ with $J/\psi\omega$, which probes the
isoscalar component of the $X(3872)$.

As for the $X(3872)\rightarrow J/\psi \pi\pi \gamma$ cascade decays,
the $J/\psi\omega$ contribution is much more important than that of
the $J/\psi\rho$, which is similar to the case of
$X(3872)\rightarrow J/\psi \pi\pi \pi$. The branching ratios of
$X(3872)\rightarrow J/\psi \pi\pi \gamma$ with the $\rho$ and
$\omega$ contribution are in order of $10^{-7}\sim10^{-6}$. However,
the contributions of the above cascade decays through the $\rho$ and
$\omega$ mesons are strongly suppressed with respect to the diagrams
which proceed either through the $\psi(2S)$ in Fig. 2(c) or the
three body decay of the $\rho$ meson in Fig. 2(d). The QED gauge
invariance demands the existence of the seagull diagram Fig. 2(d).
The branching ratio of $X(3872)\rightarrow J/\psi\rho\to J/\psi
\pi^+\pi^-\gamma$ may reach \textcolor{red}{$10^{-4}$}. The radiative transition of
$X(3872)\rightarrow J/\psi \pi^+ \pi^- \gamma$ seems to be a very
clean process to precisely study the isospin violation property of
$X(3872)$ and extract the coupling of $X(3872)$ with $J/\psi\rho$,
which probes the isovector component of the $X(3872)$.

The branching ratios of $X(3872)\rightarrow J/\psi \pi \gamma$ and
$X(3872)\rightarrow J/\psi \pi\pi \gamma$ are accessible for the
BESIII and LHCb Collaborations. With the relationships between the
branching ratios of $X(3872)\rightarrow J/\psi \pi(\pi) \gamma$ and
the coupling constants $g_{X\psi\rho/\omega}$, we can extract
$g_{X\psi\rho}$ and $g_{X\psi\omega}$ if the branching ratios of
$X(3872)\rightarrow J/\psi \pi \gamma$ and $X(3872)\rightarrow
J/\psi \pi\pi \gamma$ are measured in the near future. These
couplings encode very important information on the inner structure
of the $X(3872)$.

\section*{ACKNOWLEDGMENTS}
We are grateful to the helpful discussions with Yan-Ke Chen and
Bo-Lin Huang. This research is supported by the National Science
Foundation of China under Grants No. 11975033, No. 12070131001 and
No. 12147168. J.-Z.W. is also supported by the National Postdoctoral Program for Innovative Talent.

\end{document}